\global\def\draftcontrol{0}

   \def\versionno{ boson star stability}

\catcode`\@=11

\expandafter\ifx\csname draftcontrol\endcsname\relax\global\def\draftcontrol{0}
\fi

{\count255=\time\divide\count255 by 60
\xdef\hourmin{\number\count255}
\multiply\count255 by-60\advance\count255 by\time
\xdef\hourmin{\hourmin:\ifnum\count255<10 0\fi\the\count255}}
\def\draftdate{\number\month/\number\day/\number\year\ \ \ \hourmin }

\newcommand\makepapertitle{\par
  \begingroup
    \renewcommand\thefootnote{\@fnsymbol\c@footnote}%
    \def\@makefnmark{\rlap{\@textsuperscript{\normalfont\@thefnmark}}}%
    \long\def\@makefntext##1{\parindent 1em\noindent
            \hb@xt@1.8em{%
                \hss\@textsuperscript{\normalfont\@thefnmark}}##1}%
     \newpage
     \global\@topnum\z@   
     \@makepapertitle
     \thispagestyle{empty}\@thanks
  \endgroup
  \setcounter{footnote}{0}%
  \global\let\thanks\relax
  \global\let\makepapertitle\relax
  \global\let\@makepapertitle\relax
  \global\let\@thanks\@empty
  \global\let\@author\@empty
  \global\let\@date\@empty
  \global\let\@title\@empty
  \global\let\title\relax
  \global\let\author\relax
  \global\let\date\relax
  \global\let\and\relax
  \def\version{\let\version\@version\@gobble}
}
\def\@makepapertitle{%
  \newpage
   \ifnum\draftcontrol=1 {}
   \version\versionno
   \vskip 3em%
   \else
   \hfill\hbox to 3cm {\parbox{4cm}{\@pubnum}\hss}%
   \vskip 3em%
   \fi
   \begin{center}%
   \let \footnote \thanks
     {\LARGE {\@title}}%
     \vskip 1.5em%
     {\normalsize
       \lineskip .5em%
       \begin{tabular}[t]{c}%
         \@author
       \end{tabular}\par}%
     \vskip 1.5em%
     {\@bstract}%
     \end{center}%
     \vskip 1.5em
     \@date%
   \par
}

\gdef\@pubnum{}
\def\pubnum#1{%
  \gdef\@pubnum{#1}}

\gdef\@bstract{}
\def\Abstract#1{%
  \gdef\@bstract{%
   \parbox{\textwidth-0pc}{%
   \centerline{\bf Abstract}\penalty1000%
\kern.2cm%
\noindent
\renewcommand\baselinestretch{1.0}%
{#1}}}
}

\def\ps@paper{\let\@mkboth\@gobbletwo%
     \ifnum\draftcontrol=1
    \def\@oddfoot{\hbox to \textwidth{\tiny \versionno \hfil\tiny\draftdate}%
    \hskip -\textwidth \hbox to \textwidth{\hfil\rm\thepage\hfil}}%
     \else\def\@oddfoot{\hbox to \textwidth{\hfil\rm\thepage\hfil}}
     \fi
     \let\@evenfoot\@oddfoot
}

\def\body{\clearpage
          \pagestyle{paper}
    }

\def\@version#1{\ifnum\draftcontrol=1
\typeout{}\typeout{#1}\typeout{}
\vskip3mm\centerline{\hbox{\fbox{\normalsize{\tt DRAFT -- #1 -- }
                   {\draftdate}}}}\vskip3mm
\fi}
\let\version\@version
\long\def\eqlabel#1{\ifnum\draftcontrol=1
                    \tag@false  
                    \tag*{(\theequation) \hbox to -0.2cm{\hspace{0cm}\small{#1}\hss}}
                    \refstepcounter{equation}
                    \edef\@currentlabel{\theequation}
                    \ltx@label{#1}          
                    \else
                    \label{#1}
                    \fi
                    }
\let\st@bibitem\@bibitem
\let\st@lbibitem\@lbibitem
\ifnum\draftcontrol=1
  \def\@bibitem#1{%
    \st@bibitem{#1}\a@@label{#1}\ignorespaces}
  \def\@lbibitem[#1]#2{%
    \st@lbibitem[#1]{#2}\a@@label{#2}\ignorespaces}
  \def\a@@label#1{%
    \gdef\a@lab{\smash{\normalfont\small#1}}
    \ifvmode
      \if@inlabel
        \global\setbox\@labels\hbox{%
          \llap{\a@lab\let\a@lab\relax
                \kern\@totalleftmargin\kern\marginparsep}%
          \box\@labels}%
      \fi
    \fi}
\fi

\documentclass[12pt,letterpaper]{article}

\usepackage{amsmath,amssymb,array,calc,epsfig,rotating,bm}
\usepackage[sort]{cite}
\usepackage{graphicx}
\usepackage{psfrag,verbatim}

\ifnum\draftcontrol=1
\fi

\renewcommand\baselinestretch{1.25}
\renewcommand\section{\@startsection {section}{1}{\z@}%
                                   {-3.5ex \@plus -1ex \@minus -.2ex}%
                                   {2.3ex \@plus.2ex}%
                                   {\normalfont\large\bfseries}}
\renewcommand\subsection{\@startsection{subsection}{2}{\z@}%
                                   {-3.25ex\@plus -1ex \@minus -.2ex}%
                                   {1.5ex \@plus .2ex}%
                                   {\normalfont\normalsize\bfseries}}
\renewcommand\subsubsection{\@startsection{subsubsection}{3}{\z@}%
                                   {-3.25ex\@plus -1ex \@minus -.2ex}%
                                   {1.5ex \@plus .2ex}%
                                   {\normalfont\normalsize\it}}
\renewcommand\paragraph{\@startsection{paragraph}{4}{\z@}%
                                   {-3.25ex\@plus -1ex \@minus -.2ex}%
                                   {1.5ex \@plus .2ex}%
                                   {\normalfont\normalsize\bf}}

\numberwithin{equation}{section}

\def\revise#1       {\raisebox{-0em}{\rule{3pt}{1em}}%
                     \marginpar{\raisebox{.5em}{\vrule width3pt\
                     \vrule width0pt height 0pt depth0.5em
                     \hbox to 0cm{\hspace{0cm}{%
                     \parbox[t]{4em}{\raggedright\footnotesize{#1}}}\hss}}}}

\newcommand\nxt[1]  {\\\fnxt#1}
\newcommand{\ie}{{\it i.e.,}\ }

\def\cali         {{\cal I}}

\def\call         {{\cal L}}
\def\calm         {{\cal M}}
\def\caln         {{\cal N}}
\def\calo         {{\cal O}}

\def\del          {\partial}

\def\sqr#1#2{{\vcenter{\vbox{\hrule height.#2pt
 \hbox{\vrule width.#2pt height#1pt \kern#1pt
 \vrule width.#2pt}\hrule height.#2pt}}}}

\def\a{\alpha}
\def\b{\beta}
\def\w{\omega}

\def\dd{\delta}
\def\e{\epsilon}

\def\g{\gamma}

\def\aa1{\phi}
\def\cc1{\psi}

\def\l{\lambda}

\def\Om{\Omega}

\catcode`\@=12

\begin{document}


\title{\bf  On stability of nonthermal states in strongly coupled gauge theories}

\pubnum{%
INT-PUB-15-045}
\date{September 2, 2015}

\author{
Alex Buchel$^{1,2,4}$ and Michael Buchel$^{3}$ \\[0.4cm]
\it $^1$\,Department of Applied Mathematics, $^2$\,Department of Physics and Astronomy, \\
\it $^3$\,Faculty of Engineering\\
\it University of Western Ontario\\
\it London, Ontario N6A 5B7, Canada\\
\it $^4$\,Perimeter Institute for Theoretical Physics\\
\it Waterloo, Ontario N2J 2W9, Canada
}

\Abstract{Low-energy thermal 
equilibrium states of strongly coupled $\caln=4$ supersymmetric
Yang-Mills (SYM) theory on a three-sphere are unstable with respect to
fluctuations breaking the global $SO(6)$ R-symmetry. Using the gauge
theory/gravity correspondence, a large class of initial conditions in
the R-symmetry singlet sector of the theory was been identified that
fail to thermalize \cite{Buchel:2013uba,Balasubramanian:2014cja}.  A
toy model realization of such states is provided by {\it boson stars},
a stationary gravitational configurations supported by a complex
scalar field in $AdS_5$-gravity.  Motivated by the SYM example, we
extend the boson star toy model to include the global $SO(6)$
R-symmetry. We show that sufficient light boson stars in the
R-symmetry singlet sector are stable with respect to linearized
fluctuations.  As the mass of the boson star increases, they do suffer
tachyonic instability associated with their localization on
$S^5$. This is opposite to the behaviour of small black holes (dual to
equilibrium states of $\caln=4$ SYM) in global $AdS_5$: the latter
develop tachyonic instability as they become sufficiently light.
Based on analogy with light boson stars, we expect that the R-symmetry
singlet nonthermal states in strongly coupled gauge theories,
represented by the quasiperiodic solutions
of \cite{Balasubramanian:2014cja}, are stable with respect to
linearized fluctuations breaking the R-symmetry.
}

\makepapertitle

\body

\version\versionno
\tableofcontents

\section{Introduction and summary}\label{intro}
Consider maximally supersymmetric $\caln=4$ $SU(N)$ supersymmetric Yang-Mills theory 
on a three-sphere $S^3$ in the planar limit ($N\to \infty$, $g_{YM}^2\to 0$ 
with $g_{YM}^2 N$ kept constant) and at large 't Hooft coupling, $g_{YM}^2 N\gg 1$. 
In this limit the theory is best described by its holographic dual  \cite{m1,Aharony:1999ti} ---
type IIB supergravity in asymptotically $AdS^5\times S^5$ space-time. The 
global symmetry of the five-sphere $S^5$ geometrizes the R-symmetry of the SYM. 
To simplify the discussion, we  focus on $S^3$-invariant initial configurations of the SYM, 
and their evolution,  consistently described 
within supergravity approximation. The vacuum of the theory has a Casimir 
energy 
\begin{equation}
E_{vacuum}=\frac{3(N^2-1)}{16 L}\,,
\eqlabel{casen}
\end{equation}
where $L$ is the radius of the $S^3$. An initial state with the energy 
\begin{equation}
E=(1+\epsilon)\ \times\ E_{vacuum} > E_{vacuum}\,,
\eqlabel{adden}
\end{equation}
if it equilibrates in the future, is described by a Schwarzschild black hole in $AdS_5\times S^5$,
with the entropy $S$, the temperature $T$, and the 
size\footnote{The size is defined as a radius of the $S^3$ measured at the 
black hole horizon.} $r_+$ given by 
\begin{equation}
S(\e)=\frac{\pi N^2}{2^{3/2}} \left(\sqrt{1+\e}-1\right)^{3/2}\,,\ (T L)^2=\frac{1}{2\pi^2}\ \frac{1+\e}{\sqrt{1+\e}-1}\,, \
\left(\frac{r_+}{L}\right)^2=\frac{1}{2}
\left(\sqrt{1+\epsilon}-1\right) \,.
\eqlabel{bhn4}
\end{equation} 
The dynamical equilibration of a large class of such SYM states, with an additional assumption 
that the dynamics occurs in the R-symmetry singlet sector\footnote{The full symmetry of the dynamics is 
thus $SO(4)\times SO(6)$.}, has been discussed in  \cite{Bizon:2011gg,Buchel:2012uh}. Once a 
black hole becomes sufficiently small (light), 
\begin{equation}
\frac{r_+}{L}\ \lesssim  0.440234\qquad \Longrightarrow\qquad \epsilon\le 
\epsilon_{crit}\simeq 2.53616\,,
\eqlabel{epcrit}
\end{equation}
it suffers from the Gregory-Laflamme (GL) instability \cite{Gregory:1993vy} 
towards localization on $S^5$ \cite{Hubeny:2002xn,Dias:2015pda,Buchel:2015gxa}.
The latter implies, in particular, that {\it any} initial condition 
(off the equilibrium) of the SYM with the energy less than 
$\epsilon_{crit}$  can not equilibrate 
within the R-symmetry singlet sector; in other words, the R-symmetry must be spontaneously broken 
in the approach to thermal equilibrium for {\it any} sufficiently low-energy state. First attempts to 
construct thermally equilibrium states in $\caln=4$ SYM with broken R-symmetry were undertaken in 
\cite{Dias:2015pda}. 

Here, we point to another possibility regarding the low-energy dynamics of $\caln=4$ SYM. 
In \cite{Bizon:2011gg} (and further extended in \cite{Maliborski:2013jca}) 
it was pointed out that evolutions starting with initial data close to a single mode in AdS did not collapse. The class of 
non-collapsing solutions was  extended in \cite{Buchel:2013uba,Balasubramanian:2014cja}.
A typical representative of this class is a boson-star  \cite{Buchel:2013uba}
or a  boson-star-like \cite{Balasubramanian:2014cja} (quasi-periodic\footnote{A recent paper 
\cite{Green:2015dsa} argues that these solutions are anchors of the 
AdS-stability islands.}) 
configuration that is characterized by a broad distribution of a scalar profile in AdS 
and the dynamical evolution such that the nonlinear dispersion of the scalar energy-density 
 overcomes the focusing effects of  gravity in the AdS-cavity. 
Since the dynamics of  \cite{Buchel:2013uba,Balasubramanian:2014cja} was discussed entirely in 
$AdS_{d+1}$, it necessarily occurs in the R-symmetry singlet sector of the holographically
 dual gauge theory. It is natural to expect that unlike evolutionary trajectories that end 
up in the thermal state, the dynamics of the 
nonthermal states of \cite{Buchel:2013uba,Balasubramanian:2014cja,Bizon:2011gg,Maliborski:2013jca}  is consistently restricted 
to the R-symmetry singlet sector --- 
these states are stable with respect to localization in the compact manifold 
of the full ten-dimensional gravitational dual. The reason being the widely distributed 
energy-density profile of the scalar fields supporting the solution that shuts-off the tachyonic 
instabilities observed for small black holes in $AdS_5\times S^5$. In this paper we present
some evidence that such a scenario is indeed  realized. 
Thus, we are led to conjecture 
that there is a large class of nonthermal low-energy states in strongly coupled gauge theories 
with unbroken R-symmetry.   
 
The rest of the paper is organized as follows. In the next section we describe a toy model 
of asymptotically $AdS_5\times S^5$ holographic correspondence which supports boson stars.
We construct  boson star solutions first in the effective five-dimensional description, and further 
generalize the model to a ten-dimensional setting, where the compact manifold is a five-sphere and the 
effective five-dimensional negative cosmological constant is produced by the self-dual five-form flux.
In section  \ref{stability} we study stability of the ground state boson stars with respect to the 
linearized fluctuations breaking the $SO(6)$ symmetry of the five-sphere. We find that 
boson stars are indeed free from the tachyonic instabilities, provided they are light enough.
We emphasize that this is {\it opposite} to the fate of smeared small black holes in $AdS_5$ ---  
we expect that all states in $\caln=4$ SYM with vanishingly small energy and unbroken $R$-symmetry are non-equilibrium.  
 We conclude in section 
\ref{conclude}.
   
\section{Boson stars in a holographic toy model}

Boson stars are stationary gravitational solutions supported by a complex scalar field 
stress-energy tensor. In asymptotically AdS space-times they were originally discussed 
in \cite{Astefanesei:2003qy}.  We begin with boson stars in asymptotically $AdS_5$ space-time 
supported by a massless complex scalar field, and then extend the model  to a ten-dimensional setting. 

\subsection{Five-dimensional perspective}\label{5dsec}

Consider an effective action\footnote{We set the curvature radius of the asymptotically 
$AdS_5$ solution to $L=1$.} 
\begin{equation}
S_5=\frac{1}{16\pi G_N}\int_{\calm_5} d^5\xi \sqrt{-g}\left(R_5+12-3\del\phi\del\bar{\phi}\right)\,,
\eqlabel{5d}
\end{equation}
where $G_N$ is a five-dimensional Newton's constant, $\phi=\phi_1+i \phi_2$ is a complex scalar field, 
and 
\begin{equation}
\calm_5=\del\calm_5\times \cali\,,\qquad  \del\calm_5=R_t\times S^3\,,\qquad \cali=\left\{y\in [0,1]\right\}\,.
\eqlabel{background}
\end{equation} 
Adopting the line element as 
\begin{equation}
ds^2_5=\frac{1}{y}\left(-a e^{-2\dd}dt^2+\frac{dy^2}{4y(1-y)a}+(1-y) d\Om_3^2\right)\,,
\eqlabel{5dmetric}
\end{equation}
where $d\Om_3^2$ is the metric of unit radius $S^3$, and $a(y)$ and $\dd(y)$ are scalar functions 
of the radial coordinate $y$ describing the metric, 
and further assuming that the complex scalar field varies harmonically 
\begin{equation}
\phi_1(y,t)+i \phi_2(y,t)= p(y) e^{i \omega t}\,,
\eqlabel{sf}
\end{equation}
the equations of motion describing the boson star form the following system of ODEs:
\begin{equation}
\begin{split}
0=&p''+\frac{(2y-1)a-y+2}{y(y-1)a}\ p'-\frac{\w^2 e^{2\dd}}{4y(y-1)a^2}\ p\,,\\
0=&\dd'+2 y (y-1) (p')^2-\frac{e^{2\dd}p^2 \w^2}{2a^2}\,,\\
0=&a'+2 y(y-1)(p')^2 a+\frac{(2-y)(a-1)}{y(y-1)}-\frac{e^{2\dd}p^2\w^2}{2 a}\,.
\end{split}
\eqlabel{bseoms}
\end{equation}
A physically relevant solutions to \eqref{bseoms} must satisfy:
\nxt asymptotically at the  AdS boundary, \ie $y\to 0_+$:  
\begin{equation}
\begin{split}
p=&p_0\ y^2+\left(\frac 43 p_0-\frac{1}{12}\w^2 p_0\right)\ y^3+\calo(y^4)\,,\\
a=&1+a_2\ y^2+a_2\ y^3+\calo(y^4)\,,\\
\dd=&=2p_0\ y^4 +\calo(y^5)\,;
\end{split}
\eqlabel{uv}
\end{equation}
\nxt at the origin of AdS, \ie $z=1-y\to 0_+$:
\begin{equation}
\begin{split}
p=&p_0^h-\frac 18 p_0^h\w^2 (d_0^h)^2\ z+\calo(z^2)\,,\\
a=&1-\frac 14 (d_0^h)^2 (p_0^h)^2 \w^2\ z+\calo(z^2)\,,\\
\dd=&=\ln d_0^h-\frac 12 (d_0^h)^2 (p_0^h)^2\w^2\ z +\calo(z^2)\,.
\end{split}
\eqlabel{ir}
\end{equation}
We compute the mass $M\propto E-E_{vacuum}$  of the boson star as
\begin{equation}
\begin{split}
\end{split}
M= \int_0^1 dy\ \frac{1-y}{y^2a}\ \left(4y(1-y)a^2 (p')^2+\w^2 e^{2\dd} p^2\right)\,, 
\eqlabel{mass}
\end{equation}
and its charge $Q$  
\begin{equation}
\begin{split}
Q=&\int_{S^3}dS^3 \int_0^1 dy \sqrt{-g} J^t\,,\qquad J^\mu =i g^{\mu\nu} \left(\bar{\phi}\del_\nu \phi-\phi\del_\nu \bar{\phi}
\right)\,,\\
Q=&2\pi^2 \int_0^1 dy\ \frac{(1-y)e^\dd p^2 \w}{y^2a}\,.
\end{split}
\eqlabel{charge}
\end{equation}
Note that given $p_0$, or alternatively the charge $Q$, the numerical boson star solution is 
determined by 4 parameters,
\begin{equation}
\{\w\,,\ a_2\,,\ p_0^h\,,\ d_0^h \}\,,
\eqlabel{pars}
\end{equation}
which is precisely the order of the  system of ODEs \eqref{bseoms}.

\begin{figure}[t]
\begin{center}
\psfrag{x}{{$\frac{Q}{2\pi^2}$}}
\psfrag{y}{{$\frac{\w}{\w_j}$}}
\includegraphics[width=2.7in]{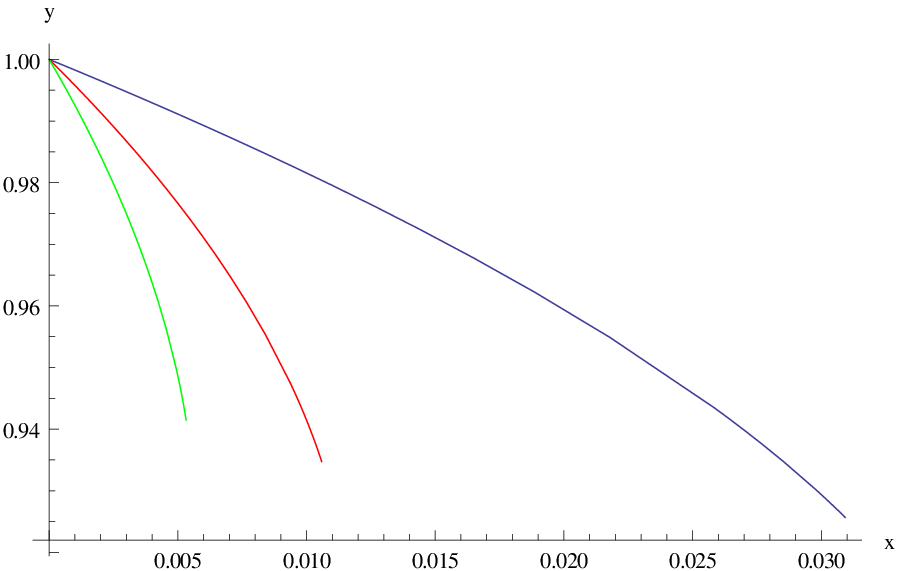}\qquad
\includegraphics[width=2.7in]{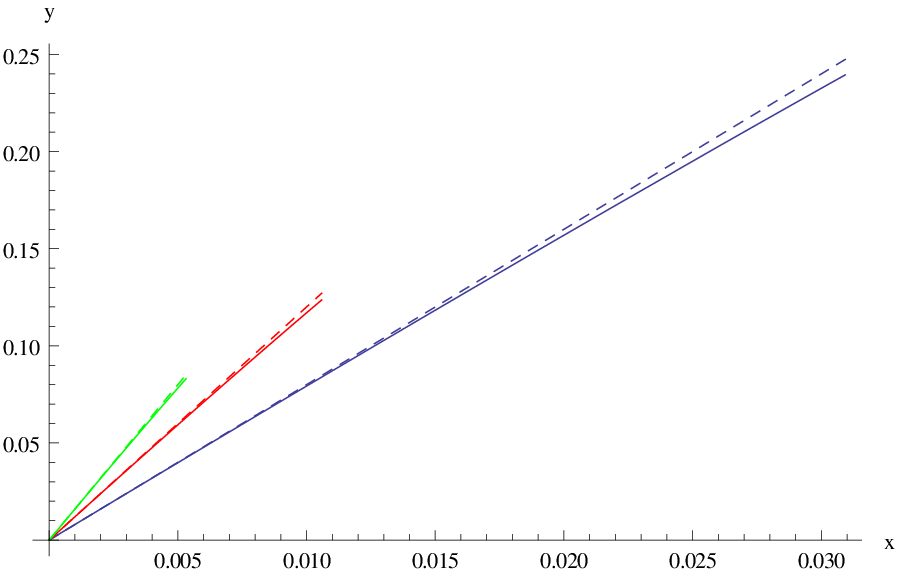}
\end{center}
  \caption{ Frequency $\w$ and mass $M$ of $j=\{0,1,2\}$ (blue,red,green) boson stars 
as a function of charge $Q$. The dashed lines represent perturbative relation between 
the mass and the charge
\eqref{pertbs}. } \label{figure1}
\end{figure}

The spectrum of light boson stars, \ie for $p_0\ll 1$, is given by
\begin{equation}
\begin{split}
&p_j=p_0 \frac{2(-1)^j}{j+2}\ y^2\ _2 F_1\left(-j\,,\, 4+j\,;\, 2\,;\, 1-y\right)+\calo(p_0^3)\,,\qquad
\w_j=4+2 j+\calo(p_0^2)\,,\\
&a_j=1+\calo(p_0^2)\,,\qquad \dd_j=\calo(p_0)^2\,, \\
&Q_j=\frac{8\pi^2p_0^2}{(j+2)^2((j+2)^2-1)} +\calo(p_0^4)\,,\  M_j=\frac{4+2j}{\pi^2}\ Q_j+\calo(Q_j^2)
=\frac{\w_j}{\pi^2}\ Q_j+\calo(Q_j^2)\,,
\end{split}
\eqlabel{pertbs}
\end{equation}
where  $j=0,1,\cdots$ is the index. 
For general $p_0$ (or $Q$) the boson stars can be found numerically, solving \eqref{bseoms}-\eqref{ir}
with a shooting method as developed in \cite{Aharony:2007vg}. The results of this analysis are presented 
in figure \ref{figure1}.

\subsection{Ten-dimensional model}\label{10d}

We would like to generalize the effective action \eqref{5d} 
to a ten-dimensional setting where we would be able to study the stability 
of the boson star solutions of \eqref{5d} with respect to linearized 
fluctuations breaking the global symmetry of the compact manifold (we choose the latter to 
be $S^5$). Our starting point is type IIB supergravity, where 
only the metric $g_{\mu\nu}^{(10)}$ and the Ramond-Ramond five-form
$F_{(5)}$ are turned on. In this case the equations of motion take the form:
\begin{equation}
G_{\mu\mu}^{(10)}\equiv  R_{\mu\nu}^{(10)}-\frac{1}{48}F_{(5)\mu\a\b\g\dd}F_{(5)\nu}^{\quad \a\b\g\dd}=0\,,\qquad dF_{(5)}=0\,,\qquad F_{(5)}=\star F_{(5)}\,.
\eqlabel{2beoms}
\end{equation}
We now add a complex scalar field $\phi=\phi_1+i \phi_2$, modifying \eqref{2beoms} as follows
\begin{equation}
\begin{split}
&G_{\mu\mu}^{(10)}\equiv R_{\mu\nu}^{(10)}-\frac{1}{48}F_{(5)\mu\a\b\g\dd}F_{(5)\nu}^{\quad \a\b\g\dd}
-3\del_\mu\phi_1\del_\nu\phi_1-3\del_\mu\phi_2\del_\nu\phi_2=0\,,\\
&dF_{(5)}=0\,,\qquad F_{(5)}=\star F_{(5)}\,.
\end{split}
\eqlabel{2beomsm}
\end{equation}
We use \eqref{2beomsm} to model ten-dimensional boson stars. The latter equations of motion are obtained 
in a toy model, where type IIB supergravity Lagrangian is supplemented with a complex scalar field:
\begin{equation}
\call_{IIB}\propto \int_{\calm_{10}} d^{10}\xi \sqrt{-g_{(10)}}\biggl(R_{(10)}+\cdots\biggr)
\ \Longrightarrow\ \int_{\calm_{10}} d^{10}\xi \sqrt{-g_{(10)}}\biggl(R_{(10)}-3\del\phi\del\bar\phi+\cdots\biggr)\,.
\eqlabel{10dlagmod}
\end{equation}   
We are interested in the most general ansatz describing solutions with $SO(4)\times SO(5)$ 
isometry\footnote{We follow here the general discussion in \cite{Buchel:2015gxa}. }. To obtain
an explicit expression for the equations determining such solutions we first
fix the reparametrization invariance such that 
\begin{equation}
g_{t\theta}=g_{x\theta}=0\,.
\eqlabel{diffeo}
\end{equation}
We can thus write the line element as\footnote{Expression for the RR five-form is the most general decomposition that preserves 
$SO(4)\times SO(5)$ symmetry of the ansatz.}, 
\begin{equation}
\begin{split}
&ds_{10}^2=-c_1^2\ (dt)^2+2g_{tx}\ dtdx+c_2^2\ (dx)^2+c_3^2\ (d\Om_3)^2 +c_4^2\ (d\theta)^2 +c_5^2\ \sin^2\theta (d\Om_4)^2\,,\cr
&F_{(5)}=(\a_0\ d\theta+\a_1\ dt+\a_2\ dx)\wedge d\Om_4+(\a_3\ d\theta\wedge dt+\a_4\ d\theta\wedge dx+\a_5\ dt\wedge dx)\wedge d\Om_3\,,\\
&c_i=c_i(t,x,\theta)\,,\qquad g_{tx}=g_{tx}(t,x,\theta)\,,\qquad \a_i=\a_i(t,x,\theta)\,,
\end{split}
\eqlabel{ansatz}
\end{equation}
where $d\Om_3$ is a volume form on a unit radius $S^3$ and  $d\Om_4$ is a volume form on a unit radius $S^4$.
Next we can eliminate $\{\a_0,\a_1,\a_2\}$ by imposing 5-form self-duality,
\begin{equation}
\begin{split}
\a_0=&-\frac{c_4 c_5^4\a_5\sin^4\theta}{c_3^3\sqrt{c_1^2c_2^2+g_{tx}^2}}\,,\
\a_1=-\frac{(\a_4c_1^2 +\a_3 g_{tx})c_5^4\sin^4\theta}{c_4c_3^3\sqrt{c_1^2c_2^2+g_{tx}^2}}\,,
\ \a_2=-\frac{(\a_3c_2^2-\a_4g_{tx}) c_5^4\sin^4\theta}{c_4c_3^3\sqrt{c_1^2c_2^2+g_{tx}^2}}\,.
\end{split}
\eqlabel{a0a1a2}
\end{equation}
The resulting equations constitute a system of partial differential equations
resulting from the eight non-trivial Einstein equations (the particular
expressions are rather involved and we will not present their explicit form at
this point):
\begin{equation}
\begin{split}
&G_{tt}^{(10)}=G_{xx}^{(10)}=G_{\Om_3\Om_3}^{(10)}=G_{\theta\theta}^{(10)}=G_{\Om_4\Om_4}^{(10)}=0\,,\qquad 
G_{tx}^{(10)}=G_{t\theta}^{(10)}=G_{x\theta}^{(10)}=0\,,
\end{split}
\eqlabel{eoms}
\end{equation}
together with the five-form  equations:
\begin{equation}
\begin{split}
0=&\del_\theta \a_3+4\cot(\theta) \a_3
-\frac{\a_5 c_4^2 c_1^2}{c_1^2c_2^2+g_{tx}^2} \del_x \ln\frac{\a_5 c_4c_5^4}{c_3^3\sqrt{c_1^2c_2^2+g_{tx}^2}}
-\a_3\biggl(\del_\theta\ln\frac{c_1c_4c_3^3}{c_2c_5^4}\\
&+\frac{g_{tx}^2}{c_1^2c_2^2+g_{tx}^2}\del_\theta\ln \frac{c_2}{c_1}\biggr)
-\frac{\a_5 g_{tx}c_4^2}{c_1^2c_2^2+g_{tx}^2}\del_t \ln\frac{\a_5 c_4c_5^4}{c_3^3\sqrt{c_1^2c_2^2+g_{tx}^2}}
-\frac{\a_4c_1^4}{c_1^2c_2^2+g_{tx}^2}\del_\theta \frac{g_{tx}}{c_1^2}\,,\\
0=&\del_\theta \a_4+4\cot(\theta) \a_4-\frac{\a_5c_2^2c_4^2}{c_1^2c_2^2+g_{tx}^2} 
\del_t \ln\frac{\a_5c_4c_5^4}{c_3^3\sqrt{c_1^2c_2^2+g_{tx}^2}}
-\a_4\biggl(\del_\theta\ln\frac{c_4c_3^3}{c_5^4}\\
&+\frac{c_1^2c_2^2}{c_1^2c_2^2+g_{tx}^2}\del_\theta\ln\frac{c_2}{c_1}\bigg)
+\frac{\a_5g_{tx}c_4^2}{c_1^2c_2^2+g_{tx}^2} 
\del_x \ln\frac{\a_5c_4c_5^4}{c_3^3\sqrt{c_1^2c_2^2+g_{tx}^2}}+\frac{\a_3c_2^4}{c_1^2c_2^2 +g_{tx}^2}\del_\theta \frac{g_{tx}}{c_2^2}\,,\\
0=&\del_t \a_3-\a_3\del_t \ln\frac{c_4c_3^3\sqrt{c_1^2c_2^2+g_{tx}^2}}{c_2^2c_5^4}
-\frac{\a_4 c_1^2}{c_2^2} \del_x \ln\frac{\a_4c_1c_5^4}{c_2c_3^3c_4}-\frac{\a_3g_{tx}}{c_2^2}\del_x\ln\frac{\a_3c_5^4}{c_3^3c_4}\\
&-\frac{\a_3g_{tx}c_1^2}{c_1^2c_2^2+g_{tx}^2}\del_x\ln\frac{g_{tx}}{c_1c_2}
+\frac{\a_4g_{tx}^2c_1^2}{c_2^2(c_1^2c_2^2+g_{tx}^2)}\del_x\ln\frac{g_{tx}}{c_1c_2}
-\frac{\a_4g_{tx}}{c_2^2}\del_t\ln\frac{\a_4c_5^4}{c_3^3c_4}\\
&-\frac{\a_4g_{tx}c_1^2}{c_1^2c_2^2+g_{tx}^2}\del_t\ln\frac{g_{tx}}{c_1c_2}\,,\\
0=&\del_\theta \a_5-\del_t \a_4+\del_x \a_3\,.
\end{split}
\eqlabel{maxwel}
\end{equation}

Boson star solutions of \eqref{eoms} and \eqref{maxwel} with
$SO(6)$ symmetry take the form:
\begin{equation}
\begin{split}
&\a_3=\a_4=0\,,\qquad \a_5=\frac{\sin^3 x e^{-\dd(x)}}{\cos^5 x}\,, \qquad 
c_1=\frac{e^{-\dd(x)}\sqrt{a(x)}}{\cos x}\,,\qquad c_2=\frac{1}{\sqrt{a(x)}\cos x}\,,\\
&g_{tx}=0\,,\qquad 
\qquad c_3=\tan x\,,\qquad c_4=c_5=1\,,\qquad \phi=p(x) e^{i\w t}\,,
\end{split}
\eqlabel{ads}
\end{equation}
where $\{p(x)\,,\, a(x)\,,\, \dd(x)\}$ are solutions of \eqref{bseoms} with the identification 
\begin{equation}
\cos^2 x=y\,.
\eqlabel{idx}
\end{equation}

\section{Stability of level-$0$ boson stars}\label{stability}

We now proceed with the stability analysis of the boson star solutions \eqref{ads} with respect to the 
linearized fluctuations partially breaking the R-symmetry to $SO(5)$. To this end we assume, to linear order in 
$\l$:
\begin{equation}
\begin{split}
&\a_3=\l  A_3(x)\ dY_\ell(\theta)\ \cos (k t)\,,\qquad \a_4=\l  A_4(x)\ dY_\ell(\theta)\ \sin (k t)\,,\\ 
&\a_5=\frac{\sin^3 x e^{-\dd(x)}}{\cos^5 x}\biggl(1+\l A_5(x)\ Y_{\ell}(\theta)\ \cos(k t)\biggr)
\\
&g_{tx}=\l f(x)\ Y_\ell(\theta)\ \sin(kt)\,,\qquad 
\ c_1=\frac{e^{-\dd(x)}\sqrt{a(x)}}{\cos x}\left(1+\l f_1(x)\ Y_{\ell}(\theta)\ \cos(k t)\right)\,,\\
&c_2=\frac{1}{\sqrt{a(x)}\cos x}\left(1+\l f_2(x)\ Y_\ell(\theta)\ \cos(k t)\right)\,,\
 c_3= \tan x\left(1+\l f_3(x)\ Y_\ell(\theta)\ \cos(k t)\right)\,,\\
&c_4=1+\l f_4(x)\ Y_\ell(\theta) \cos(k t)\,,\qquad c_5=1+\l f_5(x)\ Y_\ell(\theta)\ \cos(k t)\,,\\
&\phi=\phi_1+i \phi_2= p(x)\biggl(1+\l h_1(x)\ Y_\ell(\theta)\ \cos(kt)\biggr) 
\exp\left[ i\left(\w t+\l h_2(x)\ Y_{\ell}(\theta)\ \sin(kt)\right)\right]\,,
\end{split}
\eqlabel{anzatsfl}
\end{equation}
where $Y_\ell$ are the $S^5$-spherical harmonics,
\begin{equation}
\Delta_{S^5} Y_{\ell}\equiv -s\ Y_\ell=-\ell(\ell+4) Y_\ell \,,\qquad dY_{\ell}=\del_\theta Y_{\ell}\,.
\eqlabel{yldef}
\end{equation}
To order $\calo(\l)$, the  5-form equations \eqref{maxwel} are solved with 
\begin{equation}
A_3=A_4=A_5=0\,,\qquad f_4=f_5=0\,,\qquad f_3=-\frac 13 f_1-\frac 13 f_2\,.
\eqlabel{pass1}
\end{equation}
Next, substituting \eqref{anzatsfl} (with \eqref{pass1} and using the radial coordinate $y$, see \eqref{idx} ) into \eqref{eoms}  we find at $\calo(\l)$:
\begin{equation}
\begin{split}
&0=h_1'+\frac{p^2 e^{2\dd} \w^2 y(y-1)-4 (p')^2 a^2 y^2 (y-1)^2+2a(2 a+y-2)}{4y a^2 (y-1) p' k} p \w h_2'
\\
&-\frac{1}{48p' a^4 p y^2 (1-y)^2} \biggl(
-16 y^4 a^4(1-y)^4 (p')^4+8 a^2 y^2 (1-y)^2(p^2 e^{2\dd}\omega^2y(y-1)\\
&+2a(5a+y-2))(p')^2+4 a y (-1+y) (-2 a \omega^2 p^2-\omega^2 p^2 y+2 \omega^2 p^2+a k^2) e^{2 \dd}
\\&-\omega^4 p^4 y^2 (1-y)^2 e^{4 \dd}
-4 a^2 (a \ell^2 (y-1)+4 a \ell (y-1)+a^2+6 a y+y^2-8 a-4 y+4)
\biggr) f_2\\
&-\frac{1}{48(1-y)^{3/2} y^{3/2} a^3 p p' k} \biggl(
4 a^2 y^2(1-y)^2(6p^2e^{2\dd}\omega^2 y-a\ell^2-4 a\ell) (p')^2+a y (\ell^2 \omega^2 p^2 y\\
&-\ell^2 \omega^2 p^2+4 \ell \omega^2 p^2 y-24 a \omega^2 p^2-4 \ell \omega^2 p^2
-12 \omega^2 p^2 y+24 \omega^2 p^2+6 a k^2) e^{2 \dd}\\
&+6 p^4 y^2 \omega^4 (1-y) e^{4 \dd}+2 a^2 \ell (\ell+4) (2 a+y-2)
\biggr) f
+\frac{1}{4y a^2 (1-y) p' p} \biggl(
p^2e^{2\dd}\omega^2\\
&-4 y^2 p a^2 (1-y)^2 (p')^3+4 a^2 y (1-y)(p')^2+(p^3\omega^2 y (y-1) e^{2\dd}-2 p a(a-y+2))p'
\biggr) h_1\\
&-\frac{1}{48p' a^4 p y^2 (1-y)^2} \biggl(
-16 y^4 a^4 (1-y)^4(p')^4+(8 a^2y^3(y-1)^3\omega^2p^2e^{2\dd}\\
&+8a^3y^2(1-y)^2(a+2y-4))(p')^2
\biggr) f_1
- \frac{p h_2 e^{2\dd} k \w}{4y a^2 (y-1) p'}\,,
\end{split}
\eqlabel{eoh1}
\end{equation}
\begin{equation}
\begin{split}
&0=h_2''+\frac{2 a p' y(y-1)+ a p(2 y-1)-p y+2 p}{a (y-1) y p} h_2'
-\frac{ e^{2\dd}y(6 p^2 \w^2+ k^2)-a \ell(\ell+4)}{4(y-1) a^2 y^2} h_2\\
&+\frac{\w e^{2\dd}}{4y^{1/2} p a^3 (1-y)^{3/2}} \biggl(
-4 y^2 p a^2 (1-y)^2(p')^2+ 4 a^2 y (y-1)p'+p^3\omega^2 y (y-1)e^{2\dd}\\
&+2pa(a+y-2)
\biggr) f+\frac{h_1 e^{2\dd} k \w}{2a^2 (1-y) y}\,,
\end{split}
\eqlabel{eoh2}
\end{equation}
\begin{equation}
\begin{split}
&0=f'+\frac{(2 a y-a-2 y+4)}{2a y (y-1)} f+\frac{3 h_2 \w p^2+f_1 k}{(1-y)^{1/2} y^{3/2} a}\,,
\end{split}
\eqlabel{eof}
\end{equation}
\begin{equation}
\begin{split}
&0=f_1'+\frac{3 p^2 \w}{k} h_2'-\frac{e^{2\dd} y(6 p^2  \w^2 + k^2 )-a \ell( \ell+4)}{4(1-y)^{1/2} y^{1/2} a k} f+
\frac{1}{2a^2 (y-1) y}\biggl(-4 (p')^2 a^2(1-y)^2y^2\\
&+\omega^2 p^2y(y-1)e^{2\dd}+2a(a+y-2)\biggr) f_2 \,,
\end{split}
\eqlabel{eof1}
\end{equation}
\begin{equation}
\begin{split}
&0=f_2'+\frac{\omega^2 p^2y(y-1)e^{2\dd}+2a(2a+y-2)-4 (p')^2 a^2 y^2(1-y)^2}{4a^2 (y-1) y} f_1+\frac{y^{1/2} e^{2\dd} k f}{4(1-y)^{1/2} a}
\\
&-\frac{\omega^2 p^2y(y-1)e^{2\dd}-2a(3a-y+2)-4 (p')^2 a^2 y^2(1-y^2)}{4a^2 (y-1) y} f_2-3 p p' h_1\,.
\end{split}
\eqlabel{eof2}
\end{equation}
Additionally, there are two second order equations, which are consistent with \eqref{eoh1}-\eqref{eof2},
\begin{equation}
\begin{split}
&0=f_1''+\frac{4 (p')^2 a^2 y^2(1-y)^2-\omega^2p^2y(y-1)e^{2\dd}-2a(a+y-2)}{2a^2 (1-y) y} f_2'
+\frac{2 a y-a-y+2}{a y (y-1)} f_1'\\
&-\frac{e^{2\dd} k y^{1/2} f'}{2(1-y)^{1/2} a}
- \frac{e^{2\dd} y(6 p^2  \w^2 - k^2 )-a \ell(\ell+4)}{4(y-1) a^2 y^2} f_1
+\frac{3 p^2 e^{2\dd} \w^2 y+4 a}{2(y-1) a^2 y^2} f_2
+\frac{3e^{2\dd} p^2 k \w h_2}{2a^2 (y-1) y}\\
&- \frac{e^{2\dd} k (4 (p')^2 a^2 y^2(1-y)^2-\omega^2p^2y(y-1)e^{2\dd}-2a(2ay-y+2)) f}{8y^{1/2} (1-y)^{3/2} a^3}
+\frac{3e^{2\dd} h_1 \w^2 p^2}{2a^2 (y-1) y}\,,
\end{split}
\eqlabel{ec1}
\end{equation}
\begin{equation}
\begin{split}
&0=f_2''+\frac{e^{2\dd} k y^{1/2} f'}{2(1-y)^{1/2} a}
-\frac{4 (p')^2 a^2 y^2(1-y)^2-\omega^2p^2y(y-1)e^{2\dd}-2a(2a+y-2)}{2a^2 (y-1) y} f_1'\\
&-6 p' p h_1'+ \frac{k e^{2\dd} (4 (p')^2 a^2 y^2(1-y)^2-\omega^2p^2y(y-1)e^{2\dd}-2a(2ay-3a-y+2)) f}{8y^{1/2} (1-y)^{3/2} a^3}
\\&+\frac{2 a y-y+2}{a y (y-1)} f_2'+ \frac{e^{2\dd} k^2 y-a \ell^2-4 a \ell-8 a}{4(y-1) a^2 y^2} f_2-6 (p')^2 h_1\,.
\end{split}
\eqlabel{ec2}
\end{equation}
Further introducing 
\begin{equation}
\begin{split}
&h_1=k\ y^{\ell/2}\ H_1\,,\qquad h_2=\w\ y^{\ell/2}\ H_2\,,\qquad f=\sqrt{1-y}\ y^{(\ell+3)/2}\ F\,,\\
&f_1=\frac 1k\ y^{(\ell+4)/2}\ F_1\,,\qquad  f_1=\frac 1k\ y^{(\ell+6)/2}\ F_2\,,
\end{split}
\eqlabel{capfl}
\end{equation}
the spectrum of $SO(5)$-invariant fluctuations about $SO(6)$-symmetric boson stars is determined solving \eqref{eoh1}-\eqref{eof2}
subject to the asymptotic expansions:
\nxt asymptotically at the  AdS boundary, \ie $y\to 0_+$:  
\begin{equation}
\begin{split}
H_1=&1-\frac{6 H_{2,0}^b \w^2-3 \ell^2-\ell \w^2-8 \ell+3 k^2}{12(\ell+3)}\ y+\calo(y^2)\,,\\
H_2=&H_{2,0}^b+\frac{3 H_{2,0}^b \ell^2+H_{2,0}^b \ell \w^2+8 H_{2,0}^b \ell-3 H_{2,0}^b k^2-6 k^2}{12(\ell+3)}\ y+\calo(y^2)\,,\\
F_1=&-\frac 12 F_0^b \ell-\frac{F_0^b (\ell^3+6 \ell^2-\ell k^2+4 \ell-2 k^2)}{8(\ell+1)}\ y+\calo(y^2)\,,\\
F_2=&-\frac{F_0^b (\ell+k^2)}{2(\ell+1)}-\frac{1}{8(\ell+3) (\ell+1)}\ \biggl(8 a_2 F_0^b \ell^2+F_0^b \ell^3+F_0^b \ell^2 k^2-96 \ell p_0^2 k^2+8 a_2 F_0^b \ell\\
&+10 F_0^b \ell^2+7 F_0^b \ell k^2-F_0^b k^{4}-96 p_0^2 k^2+24 F_0^b \ell+22 F_0^b k^2\biggr)\ y+\calo(y^2)\,,\\
F=&F_0^b+\frac{(\ell^2+8 \ell-k^2+8) F_0^b}{ 4(\ell+1)}\ y+\calo(y^2)\,;
\end{split}
\eqlabel{fluv}
\end{equation}
\nxt at the origin of AdS, \ie $z=1-y\to 0_+$:
\begin{equation}
\begin{split}
H_1=&H_{1,0}^h+\biggl(-\frac14 (d_0^h)^2 H_{2,0}^h \w^2+\frac{5}{16} (d_0^h)^2 F_{1,0}^h \frac{\w^2}{k^2}-\frac18 (d_0^h)^2 H_{1,0}^h k^2+\frac18 H_{1,0}^h \ell^2+H_{1,0}^h \ell\biggr) z\\
&+\calo(z^2)\,,
\end{split}
\eqlabel{flir1}
\end{equation}
\begin{equation}
\begin{split}
H_2=&H_{2,0}^h+\biggl(-\frac34 (d_0^h)^2 H_{2,0}^h (p_0^h)^2 \w^2-\frac18 (d_0^h)^2 H_{2,0}^h k^2+\frac18 H_{2,0}^h \ell^2+H_{2,0}^h \ell\\
&-\frac14 (d_0^h)^2 H_{1,0}^h k^2\biggr) z
+\calo(z^2)\,,
\end{split}
\eqlabel{flir2}
\end{equation}
\begin{equation}
\begin{split}
F_1=&F_{1,0}^h+\biggl(\frac34 \w^2 (p_0^h)^2 (d_0^h)^2 H_{1,0}^h k^2-\frac{15}{16} F_{1,0}^h (d_0^h)^2 (p_0^h)^2 \w^2-\frac18 (d_0^h)^2 k^2 F_{1,0}^h+\frac18 F_{1,0}^h \ell^2\\
&+F_{1,0}^h \ell+\frac74
 F_{1,0}^h\biggr) z+\calo(z^2)\,,
\end{split}
\eqlabel{flir3}
\end{equation}
\begin{equation}
\begin{split}
F_2=&-\frac14 F_{1,0}^h+\biggl(-\frac14 \w^2 (p_0^h)^2 (d_0^h)^2 H_{1,0}^h k^2+\frac18 \w^2 (p_0^h)^2 (d_0^h)^2 H_{2,0}^h k^2\\&+\frac{5}{16} F_{1,0}^h (d_0^h)^2 (p_0^h)^2 \w^2
+\frac{1}{16}
 (d_0^h)^2 k^2 F_{1,0}^h-\frac{1}{48} F_{1,0}^h \ell^2-\frac{5}{24} F_{1,0}^h \ell-\frac12 F_{1,0}^h\biggr) z+\calo(z^2)\,,
\end{split}
\eqlabel{flir4}
\end{equation}
\begin{equation}
\begin{split}
F=&\frac32 H_{2,0}^h (p_0^h)^2 \w^2+\frac12 F_{1,0}^h+\biggl(\frac54 \w^2 (p_0^h)^2 H_{2,0}^h \ell+\frac{5}{12} F_{1,0}^h \ell-\frac58 H_{2,0}^h (p_0^h)^4 \w^4 (d_0^h)^2
\\&-\frac{13}{48} F_{1,0}^h (d_0^h)^2 (p_0^h)^2 \w^2-\frac{1}{24} (d_0^h)^2 k^2 F_{1,0}^h+\frac{1}{24} F_{1,0}^h \ell^2+\frac{11}{12} F_{1,0}^h+3 H_{2,0}^h (p_0^h)^2 \w^2
\\&-\frac18 \w^2 (p_0^h)^2 (d_0^h)^2 H_{2,0}^h k^2+\frac18 \w^2 (p_0^h)^2 H_{2,0}^h \ell^2-\frac14 (d_0^h)^2 H_{2,0}^h (p_0^h)^2 \w^4\biggr) z+\calo(z^2)\,.
\end{split}
\eqlabel{flir5}
\end{equation}
Note that without loss of generality we normalized linearized  fluctuations so that $H_1\bigg|_{y=0}=1$. Furthermore, the total number of parameters
characterizing the solution, 
\begin{equation}
\left\{k^2\,,\, H_{2,0}^b\,,\, F_0^b\,,\, H_{1,0}^h\,,\, H_{2,0}^h\,,\,  F_{1,0}^h\right\}\,,
\eqlabel{flpar}
\end{equation}
is precisely the order of the ODE system \eqref{eoh1}-\eqref{eof2}.

Once again, we use the shooting method developed in \cite{Aharony:2007vg}
to determine the spectrum of the linearized fluctuations. First we outline the 
computations for the light boson stars, \ie for $p_0\ll 1$, and then 
present the results for the general boson stars. We restrict out attention 
to $j=0$ (ground state) boson stars. For $j>0$ the radial profile 
of a boson star $p(y)$ has $j$ nodes inside the interval $y\in (0,1)$. 
This results in additional poles in the fluctuation equations 
\eqref{eoh1} and \eqref{eoh2} which renders our shooting method 
inapplicable\footnote{Identical technical difficulties were also 
observed in \cite{Buchel:2013uba}. These difficulties can be resolved assuming a more general fluctuation ansatz for $j>0$ as in 
\cite{Maliborski:2014rma}.}. 

\subsection{Spectrum of linearized fluctuations 
of $j=0$ light boson stars}

Using \eqref{pertbs}, to leading order in $p_0$,
\begin{equation}
p(y)\equiv p_{j=0}(y)=p_0\ y^2+\calo(p_0^3)\,,\qquad \w=\w_{j=0}=4+\calo(p_0^2) \,,
\eqlabel{leadphi}
\end{equation}
and 
\begin{equation}
\begin{split}
&H_1=H_{1,0}(y)+\calo(p_0^2)\,,\qquad H_2=H_{2,0}(y)+\calo(p_0^2)\,,\\ 
&F=p_0^2 F_{,2}(y)+\calo(p_0^4)\,,\qquad F_1=p_0^2 F_{1,2}(y)+\calo(p_0^4)\,,\qquad
 F_2=p_0^2 F_{2,2}(y)+\calo(p_0^4) \,,
\end{split}
\eqlabel{orderfl}
\end{equation}
 we find 
from \eqref{eoh1}-\eqref{eof2}:
\begin{equation}
\begin{split}
&0=H_{1,0}'+\frac{4 y  H_{2,0}'}{ k^2 (y-1)}
+\frac{(\ell(\ell+4)-k^2 y+y+3) F_{2,2}}{24 y^2 (y-1) k^2}
+\frac{(k^2-1) F_{1,2}}{24(1-y) y^2 k^2}
\\&-\frac{(2 \ell^2 y+8 \ell y+6 k^2 y) F_{,2}}{96 y^3 k^2 (1-y)}
+\frac{(\ell (y-1)-y-1) H_{1,0}}{2y (y-1)}
+\frac{(2 \ell y-2 k^2 y) H_{2,0}}{y k^2 (y-1)}\,,
\end{split}
\eqlabel{eoh1p}
\end{equation}
\begin{equation}
\begin{split}
0=&H_{2,0}''+\frac{(\ell (y-1)+5 y-3 ) H_{2,0}'}{y (y-1)}
+\frac{(\ell^2 +8 \ell -k^2 ) H_{2,0}}{4 y (y-1)}-\frac{k^2 H_{1,0}}{2y (y-1)}\,,
\end{split}
\eqlabel{eoh2p}
\end{equation}
\begin{equation}
\begin{split}
0=&F_{,2}'+\frac{(\ell y-\ell+4 y) F_{,2}}{2y (y-1)}-\frac{F_{1,2}}{y (y-1)}-\frac{48 y H_{2,0}} {y-1}\,,
\end{split}
\eqlabel{eofp}
\end{equation}
\begin{equation}
\begin{split}
0=&F_{1,2}'+48 y^2H_{2,0}'+\frac{(\ell^2-k^2 y+4 \ell) F_{,2}}{4y}
+\frac{(\ell+4) F_{1,2}}{2y}+ F_{2,2}+24 y H_{2,0} \ell \,,
\end{split}
\eqlabel{eof1p}
\end{equation}
\begin{equation}
\begin{split}
0=&F_{2,2}'+\frac{ F_{1,2}}{2y (y-1)}+\frac{(2 \ell (y-1)+10 y-2) F_{2,2}}{4y (y-1)}+\frac{k^2 F_{,2}}{4y}
-6 H_{1,0} k^2\,.
\end{split}
\eqlabel{eof2p}
\end{equation}
The system of ODEs \eqref{eoh1p}-\eqref{eof2p} can further be reduced to a single 
4th-order ODE for $H_{1,0}$:
\begin{equation}
\begin{split}
&0=H_{1,0}^{''''}+\frac{2(\ell y-\ell+7 y-4)}{y (y-1)} H_{1,0}'''+\frac{1}{2y^2 (1-y)^2} 
\biggl(3 \ell^2 y^2-5 \ell^2 y+36 \ell y^2-k^2 y^2+2 \ell^2\\
&-50 \ell y+k^2 y+94 y^2+14 \ell-106 y+24\biggr) H_{1,0}''
+\frac{1}{2y^2 (1-y)^2} \biggl(\ell^3 y-\ell^3+15 \ell^2 y-\ell k^2 y\\
&-13 \ell^2+\ell k^2+60 \ell y-5 k^2 y
-40 \ell+3 k^2+50 y-30\biggr) 
H_{1,0}'+\frac{1}{16y^2 (1-y)^2} \biggl(\ell^4+16 \ell^3\\
&-2 \ell^2 k^2+64 \ell^2-16 \ell k^2+k^4-64 k^2\biggr) H_{1,0}\,.
\end{split} 
\eqlabel{pert}
\end{equation}
Solving \eqref{pert} with the boundary conditions 
\begin{equation}
\lim_{y\to 0_+} H_{1,0}(y)=1\,,\qquad \lim_{y\to 1_-} H_{1,0}(y)={\rm finite}\,,
\eqlabel{ftxbc}
\end{equation} 
determines the perturbative spectrum $k^2$. 

Note that \eqref{pert} with \eqref{ftxbc} allows for a polynomial-in-$y$ solution:
\begin{equation}
\begin{split}
&H_{1,0}(y)=H_{1,0,\ell,n,\pm}(y)=\ _2F_1\left(-n\,,\, l+n+4\,;\, l+3\,;\, y\right)\,,\\ 
&k^2=k_{\ell,n,-}^2=(\ell+2n)^2\,,\qquad 
k^2=k_{\ell,n,+}^2=(\ell+8+2n)^2\,,
\end{split}
\eqlabel{polsolution}
\end{equation}
where $n=0,1,\cdots$ indexes the excitation level of a (light) ground state boson star;
additionally $\pm$ denotes discrete branches for a fixed $n$.
Although for different branches  $H_{1,0,\ell,n,-}=H_{1,0,\ell,n,+}$, the radial profiles 
for the remaining fluctuations do differ: {\it e.g.}, for $n=1$, solving \eqref{eoh2p}
with the appropriate boundary conditions, \eqref{fluv} and \eqref{flir2}, we find for 
$H_{2,0}=H_{2,0,\ell,n,\pm}$,
\begin{equation}
H_{2,0,\ell,1,-}=\frac{\ell+2}{4}-\frac{(\ell+5)(\ell+2)}{4(\ell+3)}\ y\,,\qquad H_{2,0,\ell,1,+}=-\frac{\ell+10}{4}+\frac{(\ell+5)(\ell+10)}{4(\ell+3)}\ y\,.
\eqlabel{n1h2}
\end{equation}

\subsection{Fluctuation spectrum of $j=0$ boson stars}

\begin{figure}[t]
\begin{center}
\psfrag{x}{{$\frac{Q}{2\pi^2}$}}
\psfrag{y}{{$k^2/k^2_{\ell,n,-}$}}
\includegraphics[width=2.7in]{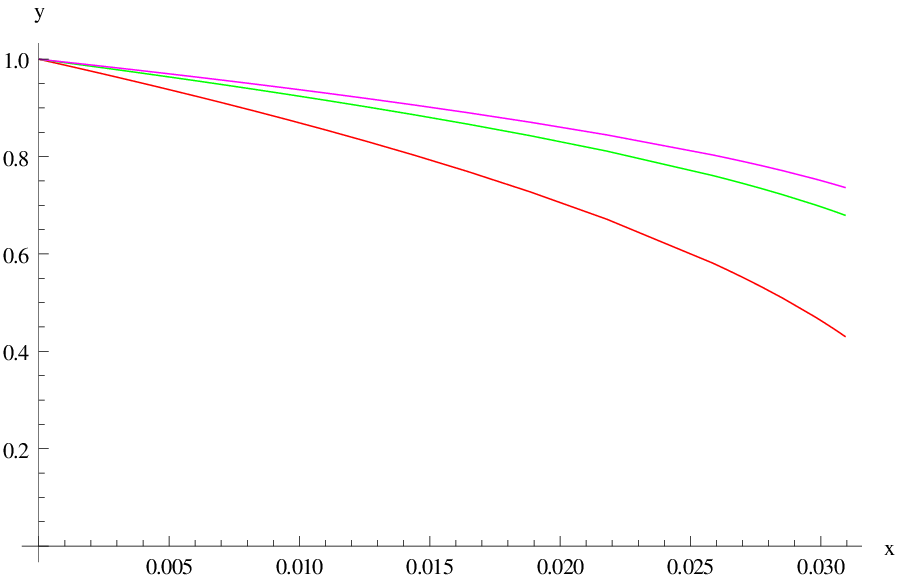}\qquad
\includegraphics[width=2.7in]{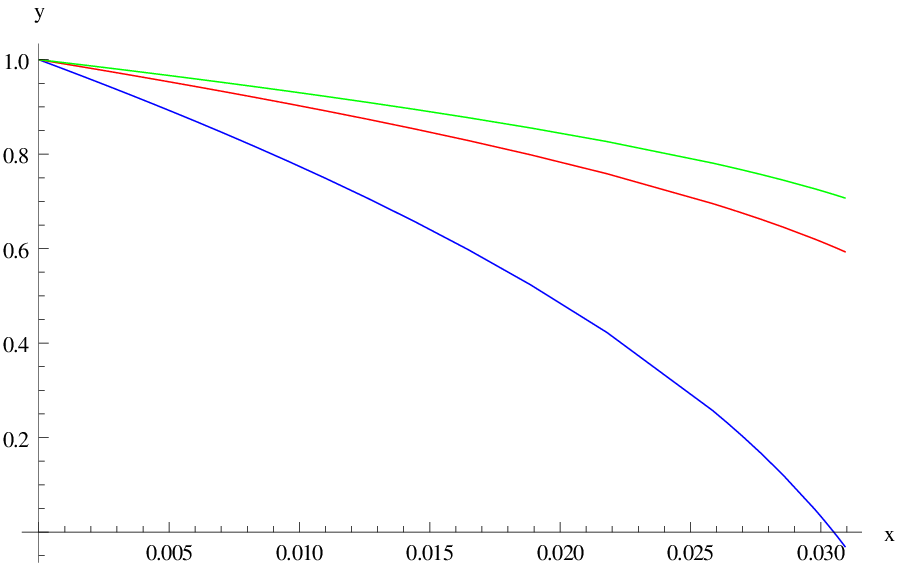}
\end{center}
  \caption{Low-lying  states in the fluctuation spectrum of $j=0$ boson 
stars as a function of charge $Q$: $\ell=\{0,1\}$ (left panel, right panel), $n=\{0,1,2,3\}$ (blue,red,green,magenta).
Leading order in $Q\to 0$ frequency eigenvalues $k^2_{\ell,n,-}$ are given by  \eqref{polsolution}.}
 \label{figure2}
\end{figure}

Perturbative in $p_0$ spectrum of the $j=0$ boson stars \eqref{polsolution} provides a starting point for the 
construction of the generic states labeled by $\{\ell,n,\pm\}$. These states are obtained numerically solving 
\eqref{eoh1}-\eqref{eof2}, subject to the asymptotic expansions \eqref{fluv}-\eqref{flir5}. The results of the 
analysis are presented in figure \ref{figure2}.
States with $\ell=0$ represent $SO(6)$-invariant fluctuations. Furthermore, the state $\{\ell=0,n=0,-\}$ is a 
zero mode corresponding to rescaling of $\l$ in \eqref{anzatsfl}. We observe that the state $\{\ell=1,n=0,-\}$
becomes tachyonic for large values of the boson star charge $Q$ (blue line, right panel). 
For small values of the charge $Q$ the frequency eigenvalues of the fluctuations are close to $k_{\ell,n,\pm}^2\, $  
(given by  \eqref{polsolution}) and thus are stable.

\section{Conclusion}\label{conclude}

The low-energy dynamics of strongly coupled gauge theories in a finite volume is rather involved. Motivated by the gauge theory/string theory correspondence 
we discussed stability of stationary solutions, {\it boson stars}, supported by a complex scalar field in $AdS_5\times S^5$. These solutions are 
typical representatives of $SO(4)$-invariant states in global $AdS_{5}$ that fail to gravitationally collapse in the limit of vanishing mass
\cite{Buchel:2013uba,Balasubramanian:2014cja,Bizon:2011gg}.  Unlike  small smeared 
black holes in $AdS_5$, which are unstable with 
respect to localization on $S^5$ 
\cite{Hubeny:2002xn,Dias:2015pda,Buchel:2015gxa}, we explicitly demonstrated 
 that boson stars are stable with respect to linearized perturbations leading to 
spontaneous breaking of the global $SO(6)$ symmetry below some critical mass 
(an in particular in the limit of vanishing mass). 
The result is far from being unexpected: the Gregory-Laflamme instability
is triggered by the localization of energy in a small region of the space-time, smeared over a large compact transverse space. 
A characteristic feature of a boson star (and also generic solutions that fail to gravitationally collapse in global $AdS_{d+1}$) 
is a broad distribution of the matter profile --- quite an opposite regime for the GL instability.
While the GL instability excludes low-energy R-symmetry singlet equilibrium states in $\caln=4$ SYM,
our analysis suggests that there is a large class of low-energy intrinsically nonthermal states in the theory, invariant under the R-symmetry. 

Our discussion was done in a toy model of the gauge theory/gravity correspondence. 
It would be interesting to study boson stars in type IIB supergravity proper. A starting point in this direction could be 
the supergravity solutions used in models of holographic superconductors, as in  \cite{Gubser:2009qm}.

~\\
\section*{Acknowledgments}
We would like to thank Pavel Kovtun for valuable discussions.
AB would like to thank Centro de Ciencias de Benasque Pedro Pascual, 
the Institute for Nuclear Theory at the University of Washington for their hospitality, and 
the Department of Energy for partial support during the completion of this work.
Research at Perimeter
Institute is supported by the Government of Canada through Industry
Canada and by the Province of Ontario through the Ministry of
Research \& Innovation. AB gratefully acknowledge further support by an
NSERC Discovery grant.

\end{document}